\documentclass[a4paper,11pt]{article}
\pdfoutput=1 

\usepackage{jheppub} 

\usepackage[T1]{fontenc} 

\definecolor{darkgreen}{rgb}{0,0.35,0}

\title{ Vacuum Degeneracy and Conformal Mass in Lovelock AdS Gravity}

\author[a]{Gabriel Arenas-Henriquez,}
\author[b]{Olivera Miskovic,}
\author[a]{Rodrigo Olea}

\affiliation[a]{Departamento de Ciencias F\'{i}sicas, Universidad Andres Bello\\ Sazi\'{e} 2212, Piso 7, Santiago, Chile.}
\affiliation[b]{Instituto de F\'isica, Pontificia Universidad Cat\'olica de Valpara\'iso \\ Casilla 4059, Valpara\'iso, Chile.}

\emailAdd{gabriel.arenas.henriquez@gmail.com}
\emailAdd{olivera.miskovic@pucv.cl}
\emailAdd{rodrigo.olea@unab.cl}

\abstract{It is shown that the notion of Conformal Mass can be defined within a given anti-de Sitter (AdS) branch of a  Lovelock gravity theory as long as the corresponding vacuum is not degenerate. Indeed, conserved charges obtained by the addition of Kounterterms to the bulk
action turn out to be proportional to the electric part of the Weyl tensor, when the fall-off of a generic solution in that AdS branch is considered. The factor of proportionality is the degeneracy condition for the vacua in the particular Lovelock AdS theory under study. This last feature explains the obstruction to define Conformal Mass in the degenerate case.
}

\begin{document}
\maketitle
\flushbottom

\section{Introduction}

\label{sec:intro}

Finite charges in a gravity theory with AdS asymptotics are a consequence of the renormalization of the action \cite{sken, deharo}.
The standard procedure to achieve this is to consider a conformal boundary at a given cutoff $z=\epsilon$ in the radial coordinate and to identify the
divergent terms that appear in both the action and its variation.
In doing so, it is particularly useful to consider the Fefferman-Graham coordinate frame \cite{fefferman}
\begin{equation}
d{s}^{2}=\frac{{\ell }^{2}}{4{z}^{2}}\,{dz}^2+\frac{{g}_{ij}(x,z)}{z^{2}}\,d{x}^{i}d{x}^{j}\,,
\end{equation}%
suitable to deal with any asymptotically AdS (AAdS) space. The near-boundary ($z=0$) expansion is written as
\begin{equation}
g_{ij}(x,z)={g_{(0)ij}}+zg_{(2)ij}+z^{2}g_{(4)ij}+\cdots\,,  \label{expansion}
\end{equation}%
where $g_{(0)ij}$ is the initial data for the problem of holographic reconstruction of the spacetime. In the context of gauge/gravity duality,
$g_{(0)ij}$ is the background metric of the boundary CFT. AdS/CFT is then realized as the matching of
boundary/bulk correlators, defined as functional derivatives with respect to the holographic source $g_{(0)ij}$.

Evaluating the gravity action on-shell, the terms that contain negative powers of $z$ will blow up at the cutoff.
These divergent pieces are a combination of the coefficients $g_{(k)ij}$ in the expansion (\ref{expansion}),  which are
not covariant from the point of view of $g_{ij}(x,z)$. Therefore, it is required to invert the series to express a given
coefficient as a series of  covariant functionals of the metric $g_{ij}(x,z)$, with an increasing number of derivatives.

The result is a renormalized AdS action in $D=d+1$ dimensions, which is supplemented by a series of intrinsic
counterterms as surface terms on top of a generalized Gibbons-Hawking term \cite{gibbons, myers}
\begin{equation}
I_{\mathrm{ren}}=I+I_{GGH}+\int\limits_{\partial \mathcal{M}}{{d}^{D-1}}x\sqrt{-h}\,{\mathcal{L}}_{\mathrm{ct}}(h,\mathcal{R},\nabla \mathcal{R})\,.
\end{equation}%

These counterterms depend on the boundary metric $h_{ij}$, the intrinsic curvature ${\mathcal{R}}_{jl}^{ik}(h)$ and its covariant derivatives.

Once the AdS gravity action has been properly renormalized, one can compute holographic 1-point functions
as functional variations with respect to the metric $g_{(0)ij}$, that is,

\begin{equation}
T^{ ij }\left[ g_{ (0) } \right] =\frac { 2 }{ \sqrt { -g_{ (0)ij } }  } \frac { \delta I_{ \mathrm{ ren } } }{ \delta g_{ (0)ij } } =\lim _{ z\rightarrow 0 }{ \left( \frac { 1 }{ { z }^{ d-2 } } { T }^{ ij } [h] \right)  }\,.
\end{equation}
In the previous relation, a suitable rescaling of the quasilocal energy-tensor \cite{brown, balasubramanian}
\begin{equation}
T^{ij}= \pi^{ij}+\frac{2}{\sqrt{-h}}\frac{\delta {\mathcal{L}}_{\mathrm{ct}}}{\delta h_{ij}}\,,
\end{equation}
and the limit to the asymptotic boundary has been taken. The tensor $\pi^{ij}$ stands for the canonical momentum associated to a radial
foliation of the spacetime.

Obtaining the counterterm series in Einstein-Gauss-Bonnet (EGB) AdS and, in general, Lovelock AdS gravity  requires the full machinery of holographic techniques. The presence of higher-order terms in the curvature, in this case, turns Holographic Renormalization into a formidable task. Computations are particularly involved, and the net outcome does not shed light on the general problem of holographic description of higher-curvature gravity theories.

On the other hand, an alternative background-independent regularization scheme for Einstein AdS gravity  was introduced in Refs.\cite{Olea2n, Olea1}. The asymptotic form of the extrinsic curvature $K_{ij}$ makes possible  the addition of a given boundary term $\mathcal{B}_{d}(h,K,{\mathcal{R}})$, which depends on the boundary metric, the extrinsic curvature and the boundary curvature
\begin{equation}
\tilde{I}_{\mathrm{ren}}={I}_{\mathrm{bulk}}+{c}_{d}\int\limits_{\partial {%
\mathcal{M}}}{{d}^{d}x\,{\mathcal{B}}_{d}}(h,K,\mathcal{R})\,,
\label{Ireg}
\end{equation}
where ${c}_{d}$ is a given coupling.
This proposal, also known as Kounterterm method, by-passes the standard holographic techniques as it is based on the addition of boundary terms linked to topological invariants and Chern-Simons densities. Despite the fact that this action manifestly spoils the Dirichlet problem for the boundary metric $h_{ij}$, its variation is both finite and expressible in terms of $\delta g_{ (0)ij} $.  In other words, $\tilde{I}_{\mathrm{ren}}$ is a renormalized Einstein AdS action, of a different sort, which allows to define holographic quantities at the boundary without the need of a quasilocal stress tensor as an intermediate step.
Holographic Renormalization is an efficient tool to cancel divergences in
the action and the corresponding correlation functions in gravity theories like Einstein or Einstein-Gauss-Bonnet
with negative cosmological constant. However, due to its perturbative nature, it is extremely complicated in
theories that contain higher powers in the curvature, as it is the case of Lovelock.

That is the reason why the extension of the Kounterterm method to Lovelock AdS gravity
\cite{kofinas,kofinas2} was a relevant step towards a holographic description of
a generic gravity theory with second-order field equations.

However, as the on-shell variation of the total action (\ref{Ireg}) does not define a Dirichlet action principle for the metric $h_{ij}$, i.e.,
\begin{equation}
\delta \tilde{I}_{\mathrm{ren}}=\int\limits_{\partial {\mathcal{M}}}d^{d}x\,\sqrt{-h}%
\left( \frac{1}{2}\,\hat{\tau}_{i}^{j}\left( h^{-1}\delta h\right)
_{j}^{i}+\Delta _{i}^{j}\,\delta K_{j}^{i}\right)\,,
\label{deltaItilde}
\end{equation}
a Brown-York quasi-local stress tensor cannot be directly read off from it. Only a {\it{tour-de-force}} approach, that is, performing asymptotic expansions in Eq.(\ref{deltaItilde}), can show which are the relevant quantities at the conformal boundary ($z \rightarrow 0$). The lack of an argument that would streamline this computation prevents from extracting holographic information from Eq.(\ref{deltaItilde}) in a straightforward way.

In a way, we would like to know how the tensor $\hat{\tau}_{i}^{j}$ is connected to a stress-energy momentum $T_{i}^{j}$ of the theory. What we know this far is that $\hat{\tau}_{i}^{j}$ provides the total energy (black hole mass plus vacuum energy) for a gravitational object in any Lovelock AdS theory as the surface integral
\begin{equation}
Q[\xi ]=\int\limits_{\Sigma _{\infty }}{{d}^{D-2}y\sqrt{\sigma }\,{u}_{j}{%
\xi }^{i}}\hat{\tau}_{i}^{j} \,,  \label{Q}
\end{equation}
where $\sigma$ is the determinant of the metric of the codimension-2 surface $\Sigma _{\infty }$ ($t=const$ and $r=\infty$), $u_{j}$ is a normal to the constant-time slice and ${\xi }^{i}$ is an asymptotic Killing vector \cite{kofinas, kofinas2}.
On the other hand, we know that the tensor $\hat{\tau}_{i}^{j}$ is separable in the form
\begin{equation}
\hat{\tau}_{i}^{j}=\tau _{i}^{j}+\tau _{(0)i}^{j}\,,
\end{equation}%
where $\tau _{(0)}$ is different from zero only in odd spacetime dimensions. This is justified by the fact that $\tau _{i}^{j}$ is a polynomial of the spacetime Riemann tensor and the metric, which can always be factorized by the AdS curvature
\begin{equation}
F_{\alpha \beta }^{\mu \nu }=R_{\alpha \beta }^{\mu \nu }
+\frac{1}{{\ell^{2}_{\mathrm{eff}} }}\,{\delta }_{\alpha \beta }^{\mu \nu }\,,
\label{F}
\end{equation}
where $\ell_{\rm{eff}}$ is an effective AdS radius, which accounts for the higher-curvature couplings in the Lovelock action, i.e., corrections respect to Einstein gravity.
Thus, for AdS vacuum ($F=0$), $\tau _{i}^{j}$ vanishes identically.
 Therefore, we know that, in general, $\tau$ and $\tau _{(0)}$ give rise to the mass of the black hole and the zero-point energy for global AdS space, respectively.

 In this paper, we relate Kounterterms charges to another definition of conserved quantities in AAdS spacetimes, which was given for Einstein gravity by Ashtekar-Magnon-Das (AMD) \cite{Ashtekar:1984zz, Ashtekar:1999jx}. This notion of energy, also known as Conformal Mass, is given in terms of the electric part of the Weyl tensor
 \begin{equation}
 \label{Eij}
 {E}_{i}^{j}=\frac{1}{D-3}\,{W}_{i\nu }^{j\mu }{n}_{\mu }{n}^{\nu }\,,
 \end{equation}
 where ${n}_{\mu }$ is a spacelike normal to the boundary, and it is given by the integral
\begin{equation}
\mathcal{H}[\xi ]=-\frac{{\ell }}{8\pi G}\int\limits_{\Sigma _{\infty }}{{d}^{D-2}y}\sqrt{\sigma }\,{u}_{j}{{E}_{i}^{j}\xi }^{i}\,.  \label{H}
\end{equation}

 The fact that there are two quantities which identically vanish for a global AdS space (the AdS curvature in Eq.(\ref{AdScurvature}) and the Weyl tensor) was the key observation to prove that ${E}_{i}^{j}$ can be obtained as a truncation of the Kounterterm charges in Einstein AdS \cite{jatkar0} and Einstein-Gauss-Bonnet AdS \cite{jatkar1} theories.

 Here, we study the generic fall-off of the metric and the curvature for black hole solutions in Lovelock AdS gravity, in order to see when it is possible to linearize the tensor $\tau _{i}^{j}$, such that it still contains the information on the energy of the system. In simple terms, this is equivalent to provide a criterion that discriminates when the Weyl tensor is able to measure the mass of Lovelock black holes with respect a global AdS background.

\section{Lovelock AdS gravity} \label{lovelock}

 Higher-dimensional gravity theories in AdS space with second-order equations of motion are of particular interest for holographic purpose. They represent a generalization of General Relativity and depend on a number parameters which provide a richer playground for gauge/gravity duality applications. At the same time, they require a minimal set of holographic data as the holographic reconstruction of the spacetime is given only in terms of the metric source at the boundary.

 The action of Lovelock gravity in a $D$-dimensional spacetime $\mathcal{M}$ with a metric $g_{\mu \nu}$,  is a series of dimensionally continued Euler densities \cite{lanczos, lovelock}
\begin{equation}
\mathcal{L}_p =\frac{1}{2^p}\,\delta _{\nu 1\cdots \nu _{2p}}^{\mu_1\cdots \mu _{2p}}\,R_{\mu _{1}\mu _{2}}^{\nu _{1}\nu _{2}}\cdots R_{\mu_{2p-1}\mu _{2p}}^{\nu _{2p-1}\nu _{2p}}\,,
\end{equation}
which are polynomials of order $p$ in the Riemann curvature ${R^\mu}_{\nu\alpha\beta}$. It is clear from the above relation that such terms exists only up to $p=N=[(D-1)/2]$. In our notation, $\delta _{\nu _{1}\cdots \nu _{2p}}^{\mu _{1}\cdots \mu _{2p}}=
\det [ \delta _{\nu _{1}}^{\mu _{1}}\delta _{\nu _{2}}^{\mu _{2}}\cdots
\delta _{\nu _{2p}}^{\mu _{2p}}] $ is completely antisymmetric product of $2p$ Kronecker deltas.

The action is, therefore, an arbitrary linear combination of $\mathcal{L}_p$,
\begin{equation}
I = \frac { 1 }{ 16\pi G } \int\limits_{ \mathcal{M} }{ d^{ D }x\sqrt { -g } \sum _{ p=0 }^{ N } \alpha _{ p }\mathcal{L}_p\,, }
\end{equation}
where $\alpha_p$ are coupling constants and $G$ is the Newton constant. In general, the set $\{\alpha_p\}$ is arbitrary, but restrictions may apply if one imposes additional conditions on the theory. In fact, in the context of gauge/gravity duality, the values of $\alpha_{p}$ are restricted by causality in a holographically dual theory \cite{camanho}.  It also been shown that, for certain values of the parameters, black holes are unstable under gravitational perturbations in EGB \cite{Zhidenko-Konoplya-EGB} and Lovelock theory \cite{Zhidenko-Konoplya-Lovelock}.

In order to see the higher-curvature terms as corrections to the Einstein-Hilbert AdS gravity, we fix the first coupling constants as $\alpha_{1}=1$ and $\alpha_{0}=-2\Lambda$, where \\
$\Lambda=-(D-1)(D-2)/2\ell^{2}$ is the cosmological constant expressed in terms of the AdS radius $\ell$. Then, the action becomes
\begin{equation}
I=\frac { 1 }{ 16\pi G } \int\limits_{\mathcal{M}}  d^{ D }x\sqrt { -g } \left( R-2\Lambda +\sum _{ p=2 }^{ N }{ \frac {\alpha_p }{ 2^{ p } } \, \delta _{ \nu _{ 1 }\cdots \nu _{ 2p } }^{ \mu _{ 1 }\cdots \mu _{ 2p } }\, R_{ \mu _{ 1 }\mu _{ 2 } }^{ \nu _{ 1 }\nu _{ 2 } }\cdots R_{ \mu _{ 2p-1 }\mu _{ 2p } }^{ \nu _{ 2p-1 }\nu _{ 2p } } }  \right)\,.
\end{equation}
The equations of motion read
\begin{eqnarray}
R_{ \nu  }^{ \mu  }-\frac { 1 }{ 2 } R{ \delta  }_{ \nu  }^{ \mu  }+\Lambda \delta _{ \nu  }^{ \mu  }-L_{ \nu  }^{ \mu  }=0\,,\label{eom}
\end{eqnarray}
where, in higher dimensions ($D>4$), the Einstein tensor with the cosmological constant acquires a contribution of higher-curvature terms through the Lanczos-Lovelock tensor
\begin{equation}
L_\mu^\nu=\sum _{ p=2 }^{ N } \frac { \alpha _{ p } }{ 2^{ p+1 } }\, \delta _{ \mu \mu _{ 1 }\cdots \mu _{ 2p } }^{ \nu \nu _{ 1 }\cdots \nu _{ 2p } }\, R_{ \nu _{ 1 }\nu _{ 2 } }^{ \mu _{ 1 }\mu _{ 2 } }\cdots R_{ \nu _{ 2p-1 }\nu _{ 2p } }^{ \mu _{ 2n-1 }\mu _{ 2p }}\,.\label{LL}
\end{equation}

Lovelock gravity possesses a plethora of different vacua. In this paper, we will focus only on AdS branches. An AdS vacuum is a maximally symmetric solution of the gravitational equations of motion (\ref{eom})-(\ref{LL})
\begin{equation}
R_{\alpha\beta}^{\mu\nu}=-\frac{1}{\ell_{\rm{eff}}^2}\,\delta_{\alpha \beta}^{\mu \nu}\,.
\end{equation}
Higher-order curvature terms change the bare AdS radius of Einstein-Hilbert gravity, $\ell$, to the effective AdS radius, $\ell_{\rm{eff}}$.

The substitution of the maximally-symmetric condition into the equations of motion gives rise to a polynomial in $\ell_{\rm{eff}}^{-2}$,
\begin{equation}
\Delta (\ell _{\mathrm{eff}}^{-2})=\sum_{p=0}^{N}{\frac{(D-3)!\,(-1)^{p-1}{\alpha }_{p}}{\left( D-2p-1\right) !}}\left( \frac{1}{\ell _{\mathrm{eff}}^{2}}\right) ^{p}=0\,, \label{Delta}
\end{equation}
such that each root of $\Delta$ defines a different sector in the theory.

\subsection{Vacuum degeneracy}

Let us consider, as an example, Einstein-Gauss-Bonnet  AdS gravity in $D\geq 5$, which is quadratic in the curvature.  We set the couplings $\alpha_2=\alpha$ in $L_\nu^\mu$ and $\alpha_p=0$ for $3\leq p\leq N$. The polynomial (\ref{Delta}) reads
\begin{equation}
\alpha (D-3)(D-4)\,\frac{1}{\ell_{\rm{eff}}^4}-\frac{1}{\ell_{\rm{eff}}^2} +\frac {1}{\ell^2} =0\,, \label{GB}
\end{equation}
leading to the effective AdS radii
\begin{equation}
\ell^{(\pm)\,2}_{\rm{eff}}=\frac{2\alpha(D-3)(D-4)}{1\pm\sqrt{1-\frac{4 \alpha(D-3)(D-4)}{\ell ^2}}}\,.
\end{equation}
EGB AdS gravity, therefore, has two \emph{simple}, different AdS vacua in the range of Gauss-Bonnet coupling $\alpha<\frac{\ell^2}{4(D-3)(D-4)}$, that is,
\begin{equation}
\ell_{\rm{eff}}^{(+)\,2}\neq \ell_{\rm{eff}}^{(-)\,2}\,. \label{nondegenerate}
\end{equation}
When $\alpha>\frac{\ell^2}{4(D-3)(D-4)}$, the AdS branch cannot be defined. At the critical point $\alpha=\frac{\ell^2}{4(D-3)(D-4)}$, there is only one \emph{degenerate} vacuum of multiplicity two with the radius
\begin{equation}
\ell_{\rm{eff}}^{(+)\,2}=\ell_{\rm{eff}}^{(-)\,2}=\frac{\ell^2}{2}\,. \label{degenerate}
\end{equation}
In five dimensions, the EGB gravity at the degenerate point in the space of parameters becomes Chern-Simons gravity, which has an enhanced gauge symmetry.
The EGB example suggests that a behavior of solutions within a degenerate branch would be different than the ones obtained as massive perturbations around a simple vacuum.

In Lovelock AdS gravity, a given vacuum with the radius $\ell_{\rm{eff}}$ is simple (or non-degenerate) if
\begin{equation}
\Delta ^{\prime }(\ell _{\mathrm{eff}}^{-2})=\sum_{p=1}^{N}\frac{(D-3)!\,(-1)^{p-1}p\alpha_{p}}{(D-2p-1)!}
\left(\frac{1}{\ell_{\mathrm{eff}}^{2}}\right)^{p-1}\neq 0\,.
\label{DeltaPrime}
\end{equation}
A vacuum with the radius $\ell_{\rm{eff}}$ is degenerate, with multiplicity $K>1$, if all derivatives of $\Delta(\ell _{\mathrm{eff}}^{-2})$ vanish up to the order $K$, i.e.,
\begin{eqnarray}
\Delta^{(q)}(\ell_{\mathrm{eff}}^{-2})&=&{\frac{1}{q!}\frac{d^q\Delta}{d(\ell_{\mathrm{eff}}^{-2})^q}}=0\,,\qquad q<K\,, \nonumber \\
\Delta^{(K)}(\ell _{\mathrm{eff}}^{-2})&=&
\sum_{p=K}^{N}\binom{p}{K}\frac{(D-3)!\,(-1)^{p-1}\alpha_{p}\,(\ell _{\mathrm{eff}}^{-2})^{p-K}}{(D-2p-1)!}\neq 0\,.
\label{DeltaK}
\end{eqnarray}
In EGB AdS gravity, it is straightforward to show that the vacua (\ref{nondegenerate}) fulfil $\Delta' \neq 0$, whereas the vacuum (\ref{degenerate}) satisfies $\Delta' = 0$ and $\Delta'' \neq 0$.

In order to get deeper insight into the role of vacuum multiplicity, in what follows we analyze the asymptotic behavior of solutions in Lovelock AdS gravity.

\subsection{Black hole solution and its asymptotic behavior}
\label{BH}

Lovelock AdS gravity possesses static black hole solutions in an arbitrary dimension $D$. The geometry, in that case, is described by the metric
\begin{equation}
{\ d }s^{ 2 }=-f(r){\ dt }^{ 2 }+\frac { 1 }{ f(r) } {\ dr }^{ 2 }+{\ r }^{2} d{\Sigma}_{D-2}^2\,,  \label{ansatz}
\end{equation}
where $d{\Sigma}_{D-2}^2=\sigma_{mn}(y)dy^mdy^n$ is the line element of the transversal section with constant curvature $k=+1,0,-1$.
The explicit form of the metric function $f(r)$ depends on the particular Lovelock theory for a  given set $\{\alpha_p\}$. It is, therefore, calculated from
the following algebraic master equation \cite{Boulware-Deser, CaiEGB, CaiLovelock, Dadhich, Camanho-Edelstein}
\begin{equation}
\sum _{p=0}^{N}\,\frac{\alpha_p(D-3)!}{(D-2p-1)!}\left(\frac{k-f(r)}{r^2}\right)^p=\frac{\mu}{r^{D-1}}\,.
\label{master}
\end{equation}
The integration constant $\mu$ is  related to the mass of the solution.

Without loss of generality, here, we deal with the case $k=1$. In particular, for global AdS space ($\mu$=0), the metric function becomes $f_{\mathrm{AdS}} =1 + \frac{r^{2}}{\ell_{\mathrm{eff}}^{2}}$, where
$\ell_{\mathrm{eff}}$ is one of the AdS vacua in Eq.(\ref{Delta}). Therefore, in asymptotically AdS space, the metric function can be written as
\begin{equation}
f(r) =1 + \frac{r^{2}}{\ell_{\mathrm{eff}}^{2}} +\epsilon (r)\,,
\end{equation}
where $\epsilon (r)$ is a function with a fast fall-off. From the above relation, it follows that the master equation (\ref{master}) in terms of the function $\epsilon (r)$ has the form
\begin{equation}
\sum_{p=0}^N\frac{(-1)^p\alpha _{p}(D-3)! }{(D-2p-1)!}\,\left(\frac{1}{\ell_{\mathrm{eff}}^2}
+\frac{\epsilon(r)}{r^2}\right)^p =\frac{\mu }{r^{D-1}}\,.
\end{equation}
This algebraic equation can be solved order by order in $\epsilon$.

Writing down the binomial expansion for small $\epsilon$,
\begin{equation}
\frac{\mu }{r^{D-1}}=\sum_{p=0}^{N}\frac{(D-3)!\,(-1)^{p}\alpha _{p}}{(D-2p-1)!}\,
\left[ 1+\frac{p\epsilon}{r^{2}}\left( \frac{1}{\ell _{\mathrm{eff}}^{2}}\right) ^{p-1}
+\frac{p(p-1)}{r^{4}}\left( \frac{1}{\ell _{\mathrm{eff}}^{2}}\right) ^{p-2}
+\mathcal{O}((\epsilon/r)^{2K})\right] \,,
\end{equation}
one recognizes the coefficients of the sum as $\Delta (\ell _{\mathrm{eff}}^{-2})$
and $\Delta ^{(q)}(\ell _{\mathrm{eff}}^{-2})$ given by Eqs.(\ref{Delta}) and (\ref{DeltaK}),
respectively. An immediate consequence of this is that the series acquires a simple form
\begin{equation}
-\frac{\mu }{r^{D-1}}=\sum_{q=1}^{N}\Delta ^{(q)}(\ell _{\mathrm{eff}}^{-2})\,\left( \frac{\epsilon }{r^{2}}\right) ^{q}\,,
\end{equation}
where it was used that the fact that $\Delta (\ell _{\mathrm{eff}}^{-2})=0$. From the criterion (\ref{DeltaK}), it is
manifest that the first non-trivial term in the
asymptotic form depends on the vacuum degeneracy condition.

For a non-degenerate vacuum, $\Delta ^{\prime }\neq 0$, we obtain
\begin{equation}
\epsilon (r)=-\frac{\mu}{\Delta^{\prime }(\ell _{\mathrm{eff}}^{-2})r^{D-3}}+\mathcal{O}(1/r^{2D-4})\,,
\end{equation}
what implies that the metric function behaves asymptotically as
\begin{equation}
f(r)=1+ \frac{r^{2}}{\ell_{\mathrm{eff}}^{2}}-\frac{\mu}{\Delta^{\prime
}(\ell _{\mathrm{eff}}^{-2})r^{D-3}}+\mathcal{O}(1/r^{2D-4})\,.
\label{solution}
\end{equation}

In turn, for a degenerate vacuum with multiplicity $K$, such as the one that defines Lovelock Unique
Vacuum gravity, the fall-off of the mass term in the metric is slower as $K$ is larger,
\begin{equation}
f(r)=k+\frac{r^{2}}{\ell _{\mathrm{eff}}^{2}}+\left( -\frac{\mu }{\Delta
^{(K)}\,r^{D-2K-1}}\right) ^{\frac{1}{K}}+\mathcal{O}(r^{-(2D-2K-2)/K})\,.
\label{f(r) K}
\end{equation}

An illustrative example of the asymptotic behavior of a solution with respect to multiplicity of the vacuum is given by the EGB AdS and Chern-Simons (CS) AdS black holes in $D=5$. For a generic solution to EGB AdS gravity, $f(r)$ behaves as
\begin{equation}
f_{\mathrm{EGB}}\sim 1+\frac{r^{2}}{{{\ell _{\mathrm{eff}}^{2}}}}-\frac{m}{r^{2}}\,,
\end{equation}
where $m$ is an effective mass parameter and the Gauss-Bonnet coupling satisfies  $\alpha>\ell^2/8$ (\ref{nondegenerate}).
At the critical point $\alpha=\ell^2 / 8$, the multiplicity of the vacuum is $K=2$ and the CS AdS black hole has a fall-off two orders slower than the EGB AdS black hole, i.e.,
\begin{equation}
f_{\mathrm{CS}}\sim 1+\frac{r^{2}}{{{\ell _{\mathrm{eff}}^{2}}}}-\nu\,,
\end{equation}
where $\nu$ is a constant.

Our goal is to derive the equivalent form of the formula of Conformal Mass for Lovelock AdS gravity. To this end, we have to impose a suitable condition on the asymptotic form of the Weyl tensor within a given branch of the theory. This can be worked out from the general fall-off of the mass term in the metric and it should turn out to be one prescribed by Ashtekar-Magnon-Das in the case of Einstein AdS gravity. From Eq.(\ref{f(r) K}), we may anticipate that the vacuum degeneracy will play an important role in the derivation of the result.

Indeed, in what follows, we show that $\Delta' \neq 0$ is a necessary condition to have a well-defined AMD mass.

\section{Weyl tensor vs. AdS curvature}

\label{weyl} In Einstein gravity, the Weyl tensor contains the modes that represent propagating
waves in vacuum. In general, Weyl tensor is the traceless part of the Riemann tensor, that is,
\begin{equation}
W_{\alpha \beta}^{ \mu \nu }=R_{ \alpha \beta }^{ \mu \nu }-
\frac { 1 }{ D-2 } \,\delta_{ [\alpha }^{ [\mu }R_{ \beta ] }^{ \nu] }
+\frac {R}{(D-1)(D-2) } \,{\delta }_{ \alpha \beta }^{\mu \nu }\,,
\end{equation}
where
\begin{equation}
 \delta_{[\alpha}^{[\mu}R_{\beta]}^{\nu]}=\delta_{\alpha}^{\mu}R_{\beta}^{\nu}-\delta_{\beta}^{\mu}R_{\alpha}^{\nu}-
\delta_{\alpha}^{\nu}R_{\beta}^{\mu}+\delta_{\beta}^{\nu}R_{\alpha}^{\mu}.
\end{equation}
 On the other hand, the only nonvanishing part of the field strength associated to the AdS group in a Riemannian theory is
\begin{equation}
F_{\alpha \beta }^{\mu \nu }=R_{\alpha \beta }^{\mu \nu }
+\frac{1}{{\ell}^{2}}\,{\delta }_{\alpha \beta }^{\mu \nu }\,. \label{AdScurvature}
\end{equation}
We will loosely refer to this quantity as AdS curvature.
A remarkable property of Einstein AdS gravity is that these two quantities  are equal on-shell.  On top of that, for global AdS space, both tensors vanish.

In Lovelock gravity, nonlinear curvature terms modify these tensors in a different manner. The AdS
curvature incorporates  the information on corrections to EH gravity in the effective AdS radius $\ell_{\mathrm{eff}}$ taking the form (\ref{F}).
In turn, the Weyl tensor evaluated on-shell, this time contains higher-order curvature terms through the Lanczos-Lovelock tensor,
\begin{equation}
W_{\alpha \beta }^{\mu \nu }=R_{\alpha \beta }^{\mu \nu }+\left( \frac{1}{\ell^2}+\frac{2L}{(D-1)(D-2)}\right)
\delta _{\alpha \beta }^{\mu \nu }-\frac{1}{D-2}\,{{\delta }_{[\alpha }^{[\mu }L_{\beta ]}^{\nu ]}}\,,
\end{equation}
where $L$ is the trace of $L^\mu_\nu$. The introduction of additional terms leads to a difference between the Weyl tensor and the AdS
curvature given by the expression
\begin{equation}
X_{\alpha\beta}^{\mu\nu}=W_{\alpha\beta}^{\mu\nu}-F_{\alpha\beta}^{\mu\nu}
=\left( \frac{1}{\ell^2}- \frac{1}{\ell^2_{\mathrm{eff}}}+\frac{2L}{(D-1)(D-2)}\right)
\delta _{\alpha \beta }^{\mu \nu }-\frac{1}{D-2}\,{{\delta }_{[\alpha }^{[\mu }L_{\beta ]}^{\nu ]}}\,,
\label{X}
\end{equation}
when field equations hold.

At this point, it remains to be seen whether the tensor $X$ contributes to the charges of the theory, defined as surface integrals at radial infinity.

To the above end, we analyze the asymptotic behavior of the higher-order curvature terms present in $X_{\alpha \beta }^{\mu \nu }$,
what can be obtained from the evaluation in the generic solution (\ref{f(r) K}). For
EH AdS  and  non-degenerate EGB AdS gravity, the fact that $X$ is subleading was shown in Refs.\cite{jatkar0} and \cite{jatkar1}.
  In Lovelock case, one should consider the generalization of Eq.(\ref{f(r) K}) which includes the
subleading contribution for large $r$,%
\begin{equation}
f(r)=k+\frac{r^{2}}{\ell _{\mathrm{eff}}^{2}}
+\left( -\frac{\mu }{\Delta^{(K)}\,r^{D-2K-1}}\right) ^{1/K}
-\frac{\Delta ^{(K+1)}}{K\Delta ^{(K)}}\left( \frac{\mu ^{2}}{\Delta ^{(K)2}\,r^{2D-2K-2}}\right)^{1/K}+\cdots \,.
\label{asymptotic f(r)}
\end{equation}

Once one knows the form of $f(r)$, it is straightforward to evaluate the asymptotic behavior of the AdS curvature, what yields
\begin{equation}
F_{\alpha \beta }^{\mu \nu }\sim \left( \frac{\mu }{\Delta ^{(K)}\,r^{D-1}}\right) ^{1/K}=\mathcal{O}\left( 1/r^{\frac{D-1}{K}}\right) \,.
\end{equation}
The above expression reduces to the known form $F_{\alpha \beta }^{\mu \nu }=\mathcal{O}\left(1/r^{D-1}\right) $ in EH and EGB theories.

Notice that the fall-off of $F_{\alpha \beta }^{\mu \nu }$ is slower
as the degeneracy $K$ increases. For example, in $D=5$, the AdS tensor
behaves as $ 1/r^4 $ in EGB AdS gravity, and as $1/r^2$ in CS AdS gravity.

As discussed above, the relevant quantity for definition of the conformal mass is the Weyl tensor. We can notice that the asymptotic form of the metric function (\ref{asymptotic f(r)}) suggests that a difference between the Weyl and AdS curvature tensors (\ref{X}) of a massive state in Lovelock AdS gravity is given as an power-expansion of the quantity $(\mu/r^{D-1})^{1/K}$. A straightforward calculation confirms this claim, showing that
\begin{equation}
X_{\alpha \beta }^{\mu \nu }\sim (K-1)\,A_K\left( \frac{\mu }{r^{D-1}}\right)^{\frac{1}{K}}+(K-2)\,B_K\left( \frac{\mu }{r^{D-1}}\right) ^{\frac{2}{K}}+\cdots\,.
\label{W-F}
\end{equation}
The coefficients $A_K$ and $B_K$ account for higher-order corrections $\alpha _{p\geq 2}$ and they identically vanish in EH gravity
($\alpha _{p\geq 2}=0$).

The above expression singles out $K=1$ as a special case with particular asymptotic behavior of the Weyl tensor. As long anticipated, for a non-degenerate vacuum ($K=1$), the difference between the Weyl tensor and the AdS curvature in any Lovelock AdS gravity is
subleading in $r$ of order $X_{\alpha \beta }^{\mu \nu }=\mathcal{O}\left(\mu ^{2}/r^{2(D-1)}\right) $. This agrees with the known result in the EGB gravity \cite{jatkar1}. When the vacuum is degenerate ($K\neq 1$), then the asymptotic decay of $X_{\alpha \beta }^{\mu \nu }$ is slower, as $\mathcal{O}\left((\mu /r^{D-1})^{1/K}\right) $.

In consequence, we  focus  on non-degenerate vacua, as the fall-off in that case is essential for the obtention of a finite mass from a charge expression linear in the Weyl tensor, as it is the matter of the discussion in the following section.

\section{Linearization of Kounterterm charges and Conformal Mass}
\label{konformalmass}
In gravity, the Noether charge associated to the isometries of spacetime provides an elegant and simple way to calculate the mass of the
black hole. Due to the fact that Noether currents are sensitive to the addition of boundary terms to the bulk action, a necessary condition
to produce the correct conserved quantities is to have a well-defined variational problem.

The prescription for the boundary term in Eq.(\ref{Ireg}) is different
 in even and odd dimensions, as shown in Refs.\cite{kofinas, kofinas2}.

In $D=2n$ dimensions, the Kounterterm series reads
\begin{equation}
\mathcal{B}_{2n-1}=2n\sqrt {-h} \int\limits_{0}^{1} du\,
\delta_{j_1\dots j_{ 2n-1 } }^{i_1\dots i_{2n-1}}K_{i_1}^{j_1}
\left( \frac{1}{2} \mathcal{R}_{i_2 i_3}^{j_2 j_3}-u^2  K_{i_2}^{j_2} K_{i_3}^{j_3} \right) \dots
\left( \frac{1}{2} \mathcal{R}_{i_{2n-2} i_{2n-1}}^{j_{2n-2} j_{2n-1}}-u^2  K_{i_{2n-2}}^{j_{2n-2}} K_{i_{2n-1}}^{j_{2n-1}} \right) ,
\label{evenk}
\end{equation}
where the corresponding coupling constant is
\begin{equation}
c_{2n-1}=-\frac {1}{16\pi nG}\sum _{p=1}^{n-1}
\frac{p\alpha_p}{(D-2p)!}{(-\ell_{\mathrm{eff}}^2)^{n-p}}\,.
\end{equation}

In $D=2n+1$ dimensions, the boundary term takes the form
\begin{eqnarray}
\mathcal{B}_{2n}=2n\sqrt{-h}\int\limits_{0}^{1}du\int\limits_{0}^{u}ds\,
\delta_{j_1\dots j_{2n-1}}^{i_1\dots i_{2n-1}} K_{i_1}^{j_1}
\left(\frac{1}{2}\,\mathcal{R}_{i_2 i_3}^{j_2j_3}-u^2K_{i_2}^{j_2}
K_{i_3}^{j_3}+\frac{s^2}{\ell_{\mathrm{eff}}^2}\,\delta_{i_2}^{j_2}
\delta_{i_{ 3 } }^{j_{ 3 } } \right) \times \dots  \nonumber \\
\dots \times \left(\frac{1}{2}\,\mathcal{R}_{i_{2n-2}{i}_{2n-1}}^{j_{2n-2}j_{2n-1}}
-u^2 K_{i_{2n-2}}^{j_{2n-2}}
K_{i_{2n-1}}^{j_{2n-1}}+\frac{s^2}{\ell_{\mathrm{eff}}^2}
\delta_{i_{2n-2}}^{j_{2n-2}}\,\delta_{i_{2n-1}}^{j_{2n-1}}\right),
\end{eqnarray}
with the coupling constant
\begin{equation}
c_{ 2n }=-\frac { 1 }{ 16\pi n G } \left[ \int\limits_{ 0 }^{ 1 } du (1-u^2)^{n-1}  \right]^{ -1 }
\sum _{ p=1 }^{ n }%
\frac{ p\alpha_{ p } }{( D-2p)! } \,(-\ell_{\mathrm{eff}}^2)^{n-p}\,.
\label{c_odd}
\end{equation}
The use of a parametric integrations in $u$ and $s$ is not only a mere formality, but allows to write
down the whole series in a very compact form, as well. It also enormously simplifies the derivation of
Noether charges and, in particular, it will be extensively used in the derivation of Conformal Mass for
the theory.
The above boundary terms depend explicitly on $\ell_{\mathrm{{eff} }}$. In a way, that means that the information on the vacuum
of the theory and its multiplicity is encoded both in $B_{d}$ and the corresponding couplings $c_{d}$.

At this point, we motivate the main result reported below by the prospect of AMD mass being generalized to Lovelock AdS gravity.

As the concept of Conformal Mass is not necessarily linked to the addition of boundary terms to the gravity action, it is
not guaranteed that its applicability can be extended to this case.

The first evidence in that direction comes from the fact that, in both even and odd dimensions, the charge density tensor $\tau_i^j$ is a polynomial in the curvature that can be always factorized by the AdS curvature $F^{ij}_{kl}$ (with boundary indices). This implies that the black hole charge vanishes identically for global AdS spacetime \cite{kofinas,kofinas2}.

In what follows, we work out a formula for Conformal Mass in Lovelock AdS gravity, which appears as a truncation of
the Kounterterm charges to the linear order in the curvature.

As we shall see below, when the AdS vacuum in non-degenerate, the polynomial that multiplies the AdS curvature is at most finite. Therefore, it only contributes with a proportionality factor. This is the key ingredient to transform the Noether charge in a formula proportional to the Weyl tensor. On the contrary, in the degenerate vacuum case, the expression turns nonlinear and cannot be truncated.

 This explicit derivation of the AMD formula for Lovelock theory depends on whether the dimension is odd or even. In next subsections, we highlight the main  points in the linearization of the charge in both cases. Further details are provided in Appendix \ref{truncationapp}.

\subsection{Even dimensions}
\label{even_dim}

In $D=2n$ dimensions, the vacuum energy vanishes, i.e., $\tau^{j}_{(0)i}=0$. Thus, the only contribution to the Noether charge comes from the
charge density tensor $\tau_{i}^{j}$, whose explicit form is
\begin{eqnarray}
\tau_{i}^{j} &=&\frac{{{\ell _{\mathrm{eff}}^{2n-2}}}}{16\pi G}\,\delta
_{i_{1}\cdots i_{2n-1}}^{jj_{2}\cdots
j_{2n-1}}\,K_{i}^{i_{1}}\sum_{p=1}^{n-1}{\frac{p{\alpha }_{p}}{{2}%
^{n-2}\left( 2n-2p\right) !}}\left( \frac{1}{{{\ell _{\mathrm{eff}}^{2}}}}%
\right) ^{p-1}R_{j_{2}j_{3}}^{i_{2}i_{3}}\cdots
R_{j_{2p-2}j_{2p-1}}^{i_{2p-2}i_{2p-1}}\times  \nonumber \\
&&\times \left[ {\left( \frac{1}{{{\ell _{\mathrm{eff}}^{2}}}}\right)
^{n-p}\delta _{j_{2p}j_{2p+1}}^{i_{2p}i_{2p+1}}}\cdots \delta
_{j_{2n-2}j_{2n-1}}^{i_{2n-2}i_{2n-1}}-{(-1)}%
^{n-p}R_{j_{2p}j_{2p+1}}^{i_{2p}i_{2p+1}}\cdots
R_{j_{2n-2}j_{2n-1}}^{i_{2n-2}i_{2n-1}}\right] \,.  \label{qeven}
\end{eqnarray}
Then, the charge density tensor can be factorized by the AdS curvature as
\begin{equation}
\tau_{i}^{j}=\frac{{{\ell _{\mathrm{eff}}^{2n-2}}}}{16\pi G}\,{\delta }_{{i}%
_{1}{i}_{2}\dots {i}_{2n-1}}^{j{j}_{2}\dots {j}_{2n-1}}{K}_{i}^{{i}_{1}}
F_{{j}_{2}{j}_{3}}^{{i}_{2}{i}_{3}}\mathcal{P}_{j_4\dots j_{2n-1}}^{i_4\dots i_{2n-1}}(R)\,,  \label{q}
\end{equation}
where the form of the polynomial $\mathcal{P}(R)$, of order $n-2$ in the curvature, is given by Eq.(\ref{peven}).

The crucial step is to evaluate the charge density (\ref{q}) in the asymptotic region, and to use a
power-counting argument in the radial coordinate $r$, in order to see which are the terms that do contribute
in the limit $r\rightarrow\infty$. In that way, we will obtain a consistent truncation of the charge, which still is
able to give rise to the black hole mass in Lovelock gravity.

As argued in the previous section, the asymptotic behavior of the fields depends on degeneracy of the vacuum. A non-degenerate vacuum
has the fall-off of the Weyl tensor $W \sim \mathcal{O}(1/r^{D-1})$. This was a key ingredient in the definition of Conformal Mass for AAdS spaces in Einstein theory \cite{Ashtekar:1984zz, Ashtekar:1999jx}. In the non-degenerate case, Eq.(\ref{W-F}) implies that a difference between the Weyl and AdS tensor is subleading,
$F=W+\mathcal{O}(1/r^{2D-2})$. It is also known that the corresponding expansions of the bulk curvature and extrinsic curvature are given by
\begin{eqnarray}
K_{j}^{i} &=&-\frac{1}{\ell _{\mathrm{eff}}}\,\delta _{j}^{i}+\mathcal{O}(1/r)\,,\nonumber \\
R_{kl}^{ij} &=&-\frac{1}{\ell _{\mathrm{eff}}^{2}}\,\delta _{kl}^{ij}+\mathcal{O}(1/r^{D-1})\,,
\label{KR}
\end{eqnarray}
yielding a finite value for $\mathcal{P}$, as well. It is not difficult to see that, in order to have a
finite charge $Q[\xi]$, the charge density has to be $\tau \sim \mathcal{O}(1/r^{D-1})$, which is exactly the order of $W$.
In turn, all the other quantities contribute --at most-- to the finite order.

This way of thinking has a concrete realization when one replaces the extrinsic curvature and the bulk curvature in the charge density tensor (\ref{q}) by its leading-order (finite) part, what produces
\begin{eqnarray}
\tau_i^j =-\frac{{\ell _{\mathrm{eff}}^{2n-3}}}{16\pi G}\,{\delta }_{{ii}_{2}
\cdots {i}_{2n-1}}^{j{j}_{2}\cdots {j}_{2n-1}}\,{{W}_{{j}_{2}{j}_{3}}^{{i}_{2}{i}_{3}}}
\mathcal{P}_{j_4\cdots j_{2n-1}}^{i_4\cdots i_{2n-1}}\left( -\ell _{\mathrm{eff}}^{-2}\delta^{[2]} \right) +\mathcal{O}(1/r^{2D-2})\,.
\label{q_expanded}
\end{eqnarray}
The explicit evaluation of the polynomial leads to
\begin{eqnarray}
\mathcal{P}_{j_4\dots j_{2n-1}}^{i_4\dots i_{2n-1}}\left({{-\ell _{\mathrm{eff}}^{-2}}\delta^{[2]}} \right) &=&
\frac{\Delta'(\ell _{\mathrm{eff}}^{-2})}{2^{n-1}\ell_{\mathrm{eff}}^{2(n-2)}(2n-3)!}\,
\delta_{j_4j_5}^{i_4j_5}\cdots\delta_{j_{2n-2}j_{2n-1}}^{i_{2n-2}j_{2n-1}}\,,
\end{eqnarray}
where we have identified the sum as the degeneracy condition (\ref{DeltaPrime}). It is clear that the expression is non-vanishing only if the vacuum is a simple root, i.e., $\Delta'\neq 0$.  Plugging in $\mathcal{P}$ in the charge density tensor (\ref{q_expanded}) and contracting the deltas, one gets
\begin{equation}
\tau_i^j=-\frac{\ell _{\mathrm{eff}}}{16\pi G}\,\frac{\Delta'(\ell _{\mathrm{eff}}^{-2})}{2(2n-3)}\,
\delta_{ii_2i_3}^{jj_2j_3}\,{W_{j_2j_3}^{i_2i_3}}+{\mathcal{O}}(1/r^{2D-2})\,.
\end{equation}

In order to have the Noether charge properly written as a formula for Conformal Mass, $\tau_i^j$ must be proportional to the electric part of the Weyl tensor.
Not surprisingly, this is the case. Expanding the contractions of the Weyl tensor,
\begin{eqnarray}
\delta _{ii_{2}i_{3}}^{jj_{2}j_{3}}\,W_{j_{2}j_{3}}^{i_{2}i_{3}}
&=&2\left(\delta _{i}^{j}W_{kl}^{kl}-2W_{ki}^{kj}\right)\,,
\end{eqnarray}
and using  the fact that the Weyl tensor is traceless, we can show that
\begin{equation}
W_{kl}^{kl} =0\,,\qquad W_{\mu i}^{\mu j}=W_{ri}^{rj}+W_{ki}^{kj} =0\,.
\end{equation}
In doing so, we obtain
\begin{eqnarray}
\delta _{ii_{2}i_{3}}^{jj_{2}j_{3}}\,W_{j_{2}j_{3}}^{i_{2}i_{3}}
&=&-4W_{ki}^{kj}=4W_{ri}^{rj}\,.  \label{deltaw}
\end{eqnarray}
By definition, the electric part of the Weyl tensor in even dimensions reads
\begin{equation}
E_i^j=\frac{1}{2n-3}\,{W}_{ri}^{rj}\,,
\end{equation}
where we have used the normal to the boundary $n_{\mu }=(n_r,n_i)=(N,0)$ (Appendix \ref{gauss}) in the general relation Eq.(\ref{Eij}).
Then, as a consequence of dropping subleading contributions in the curvature, we obtain
\begin{eqnarray}
\tau_i^j &=&-\frac{{\ell _{\mathrm{eff}}}}{8\pi G}\,\Delta'(\ell _{\mathrm{eff}}^{-2})\,E_{i}^{j}\,.
\label{q=E}
\end{eqnarray}

\subsection{Odd dimensions}

In a similar fashion as in even-dimensional case, in odd dimensions $D=2n+1$, the mass for black hole solutions to a generic Lovelock AdS gravity theory can be attributed to the quantity $\tau_i^j$,
\begin{eqnarray}
\tau_{i}^{j} &=&\frac{1}{16\pi G2^{n-2}}\,\delta_{i_{1}\cdots i_{2n}}^{jj_{2}\cdots j_{2n}}\,K_{i}^{i_{1}}\delta_{j_{2}}^{i_{2}}\times \nonumber\\
&&\hspace{-1cm}\times \left[ \sum_{p=1}^{n}\frac{p\alpha _{p}}{(2n-2p+1)!}\,R_{j_{3}j_{4}}^{i_{3}i_{4}}\cdots
R_{j_{2p-1}j_{2p}}^{i_{2p-1}i_{2p}}\,\delta_{j_{2p+1}j_{2p+2}}^{i_{2p+1}i_{2p+2}}\cdots \delta_{j_{2n-1}j_{2n}}^{i_{2n-1}i_{2n}}\right. +\nonumber\\
&&\hspace{-1cm}+\left. 16\pi G\,nc_{2n}
\int\limits_{0}^{1}du\left( R_{j_{3}j_{4}}^{i_{3}i_{4}}+\frac{u^{2}}{{{\ell _{\mathrm{eff}}^{2}}}}\,\delta _{j_{3}j_{4}}^{i_{3}i_{4}}\right)
\cdots \left( R_{j_{2n-1}j_{2n}}^{i_{2n-1}i_{2n}}+\frac{u^{2}}{{{\ell _{\mathrm{eff}}^{2}}}}\,\delta _{j_{2n-1}j_{2n}}^{i_{2n-1}i_{2n}}\right) \right]
\label{qodd}
\end{eqnarray}
with the coupling constant given by Eq.(\ref{c_odd}).

The above expression can be factorized by the AdS curvature
\begin{equation}
\tau_{i}^{j}=\frac{{{\ell _{\mathrm{eff}}^{2n-2}}}}{16\pi G}\,{\delta }_{{i}%
_{1}{i}_{2}\dots {i}_{2n-1}}^{j{j}_{2}\dots {j}_{2n-1}}{K}_{i}^{{i}_{1}}
F_{{j}_{2}{j}_{3}}^{{i}_{2}{i}_{3}}\tilde{\mathcal{P}}_{j_4\dots j_{2n-1}}^{i_4\dots i_{2n-1}}(R)\,,
\end{equation}
where $\tilde{\mathcal{P}}(R)$ is a polynomial  of degree $n-2$ in the Riemann tensor, as shown in Eq.(\ref{tildeP}).
 The details of this computation are given, to a certain extent, in Appendix (\ref{oddapp}).
 Then, one can follow a strategy that mimic the procedure in the even-dimensional case. As a matter of fact, for a non-degenerate  vacuum,  $F$ behaves as $W+\mathcal{O}(1/r^{2D-2})$ where $W={\mathcal{O}}(1/r^{D-1})$, such that the extrinsic and intrinsic curvatures are finite, as dictated by Eq.(\ref{KR}).
 Therefore, the charge integrand takes the form
\begin{equation}
\tau_{i}^{j}=-\frac{{{\ell _{\mathrm{eff}}^{2n-3}}}}{16\pi G}\,{\delta }_{{i}%
{i}_{2}\dots {i}_{2n-1}}^{j{j}_{2}\dots {j}_{2n-1}}
W_{{j}_{2}{j}_{3}}^{{i}_{2}{i}_{3}}\tilde{\mathcal{P}}_{j_4\dots j_{2n-1}}^{i_4\dots i_{2n-1}}(-\ell_{\mathrm{eff}}^{-2}\,\delta^{[2]})\,.
\end{equation}

A straightforward calculation reveals that $\tilde{\mathcal{P}}(-\ell^{-2}_{\mathrm{eff}}\delta^{[2]})$ is proportional to
the degeneracy condition, in such a way that
\begin{equation}
\tau_{i}^{j}=-\frac{{\ell _{\mathrm{eff}}}}{16\pi G}\,\frac{\Delta ^{\prime
}(\ell _{\mathrm{eff}}^{-2})}{2\left( 2n-2\right) }\,{\delta }_{{ii}_{2}{i}_{3}}^{j{j}_{2}j_{3}}\,{{W}_{{j}_{2}{j}_{3}}^{{i}_{2}{i}_{3}}}\,.
\end{equation}
 Notice that the subleading terms were conveniently  dropped out again. This means that, even though the charge expression appears as truncated, it stills provides the same answer for the mass of a black hole within that branch of the theory. Hence, the linearized charge density tensor and the electric part of the Weyl tensor are related by the degeneracy condition, that is,
\begin{eqnarray}
\tau_{i}^{j} &=&-\frac{{\ell _{\mathrm{eff}}}}{8\pi G}\,\Delta'(\ell _{\mathrm{eff}}^{-2})\,E_{i}^{j}\,.
\end{eqnarray}

Thus, we have shown that both in even and odd dimensions, the linearization of  Kounterterm charges  leads to  a generalization
of the AMD mass $\mathcal{H}[\xi ]$ (\ref{H}) to Lovelock AdS gravity
\begin{equation}
\mathcal{H}_{\mathrm{Lovelock}}[\xi ]=-\frac{{\ell _{\mathrm{eff}}}}{8\pi G}\,\Delta ^{\prime }(\ell _{%
\mathrm{eff}}^{-2})\int\limits_{\Sigma _{\infty }}{{d}^{2n-2}y}\sqrt{\sigma }\,{u}_{j}{%
{E}_{i}^{j}\xi }^{i}\,.  \label{H_Lovelock}
\end{equation}

From the above formula, it is evident that, for a degenerate vacuum of the theory ($\left.\Delta'=0\right.$), Conformal
Mass  vanishes identically.
This definition of energy for massive gravitational objects reproduces known results for particular cases.

In fact, in EH AdS gravity, switching off all higher-curvature corrections, we have $\Delta'(\ell _{\mathrm{eff}}^{-2})=\alpha_{1}=1$
and $\ell_{\mathrm{eff}}=\ell$. The corresponding charge is the one originally derived by
Ashtekar-Magnon-Das \cite{Ashtekar:1984zz, Ashtekar:1999jx}.
In EGB AdS case, the constant of proportionality is
\begin{equation}
\Delta'(\ell _{\mathrm{eff}}^{-2}) =\alpha _1-\frac{2\alpha_2}{\ell _{\mathrm{eff}}^2}\,(D-3)(D-4)\,,
\label{deltaegb}
\end{equation}
such that the charge reads
\begin{equation}
\mathcal{H}_{\mathrm{EGB}}[\xi ] =-\frac{{\ell _{\mathrm{eff}}}}{8\pi G}\,\left[ 1-\frac{2\alpha }{\ell _{\mathrm{eff}}^2}\,(D-3)(D-4) \right]
\int\limits_{\Sigma _{\infty } }d^{2n-2}y\,\sqrt{\sigma }\,u_j\,E_i^j\,\xi^i\,,  \label{cmegb}
\end{equation}
as shown previously in Ref.\cite{jatkar1}. The expression (\ref{deltaegb}) is zero when evaluated at the critical point, where the vacuum has
multiplicity $K=2$. This fact implies that $\mathcal{H}_{\mathrm{EGB}}[\xi ]$ vanishes, as well.

\section{Conclusions}
In this work, we have extended the concept of Conformal Mass to any branch of Lovelock AdS gravity, as long as the corresponding vacuum  is non-degenerate. The AMD energy appears as a consistent truncation,  to the linear order in the curvature, of the asymptotic charges associated to a gravity action renormalized by the addition of Kounterterms.

The fact that the resulting AMD formula for Lovelock gravity is proportional to the degeneracy condition reflects an obstruction to the linearization of the theory.  As a consequence, for a degenerate sector, the information on the black hole mass is carried to the boundary by the tensor $\tau_{i}^{j}$, which is no longer linear in the curvature. This is justified by the slower falloff of the AdS curvature in that case. A finite conserved charge, defined as a surface integral at radial infinity, would require additional Riemann tensors in it, in order to produce an asymptotic behavior  $\tau_{i}^{j}=\mathcal{O}(1/r^{D-1})$.
The limiting case in the above picture is given by $K=n$, in odd dimensions ($D=2n+1$). In that situation, the theory corresponds to Chern-Simons gravity with an AdS vacuum of maximal degeneracy such that $\tau_{i}^{j}$ vanishes, in a nontrivial way \cite{Fan-Cheng-Lu}. For that particular point in the parameter space (Lovelock couplings), the energy for dimensionally-continued black holes comes from a formula proportional to $\tau_{(0)i}^{j}$, what in other cases is associated to the vacuum energy \cite{CS-MOTZ,MO-DCG}.

 Although,  in the present work, our interest focuses on comparing two different notions of conserved quantities for AAdS gravity, our ultimate goal is to work out holographic quantities at the boundary of  Lovelock theory.

In that respect, we can get some insight from the example of Einstein AdS gravity in five dimensions, where the quasilocal stress tensor is separable as \cite{Ashtekar:1999jx,Hollands-Ishibashi-Marolf}

\begin{equation}
T_{i}^{j}=E_{i}^{j}+\Delta_{i}^{j}\,,
\label{Tseparable}
\end{equation}%
such that the trace of $\Delta_{i}^{j}$ gives rise to the Weyl anomaly, upon an appropriate rescaling in the normal coordinate to the boundary. At the same time, $\Delta_{i}^{j}$ is responsible for a nonzero vacuum energy for global AdS space.

The generalization of this result to Lovelock AdS gravity passes by understanding the role of the degeneracy condition of order $K\neq 1$ in the definition of a tensor $\tilde{E}_{i}^{j}$ which shares similar properties with $E_{i}^{j}$. Indeed, $\tilde{E}_{i}^{j}$ should be traceless but nonlinear in the curvature of the spacetime. This would allow to readily identify the conformal anomaly from a relation similar to Eq.(\ref{Tseparable}).
 All of above goes along the line the argument that standard holographic techniques break down in a degenerate Lovelock theory. As a matter of fact, in the maximally-degenerate case in odd dimensions (Chern-Simons AdS gravity) none of the coefficients $g_{(k)ij}$ with $k=2,..,2n-2$ can be determined in terms of $g_{(0)ij}$.  Therefore, the corresponding holographic description is rather unusual \cite{Banados-Olea-Theisen,Grozdanov-Starinets}. The fact that the form of Kounterterm series is universal, regardless the particular Lovelock AdS theory under study, may be of great help to deal with this issue. We hope to report on this point elsewhere.

\acknowledgments
We thank G. Anastasiou and G. Kofinas for helpful discussions. This work was supported in part by the Grants FONDECYT No.1131075 and 1170765,
VRIEA-PUCV No. 039.428/2017 and No.123.752/2017, UNAB Grant DI-1336-16/RG and CONICYT Grant DPI 20140115. G.A.H. is a UNAB M.Sc. Scholarship holder.

\appendix

\section{Gauss-normal foliation}
\label{gauss}

Any line element for a given spacetime $\mathcal{M}$ can be cast in Gaussian coordinates
\begin{equation}
ds^2=N^2(r)\,dr^2+h_{ij}(r,x)\,dx^idx^j\,,
\end{equation}
where $h_{ij}$ is the induced metric  at a fixed $r$.
This is particularly useful to express bulk quantities in terms of boundary tensors.
In this frame, the extrinsic curvature is defined by the formula
\begin{equation}
K_{ij}=-\frac{1}{2}\mathcal{L}_nh_{ij}=-\frac{1}{2N} {\partial_r }h_{ij}\,,
\end{equation}
where $\mathcal{L}_n$ is the Lie derivative along a radial normal $n_{\mu}=N\delta_{\mu}^{r}$. The
foliation leads to the Gauss-Codazzi relations
\begin{eqnarray}
R_{jl}^{ir}&=&\frac { 1 }{ N }\, ( {\nabla }_{ l }{\ K }_{ j }^{i }-{\nabla }_{ j }{K }_{ l }^{ i })\,,  \nonumber \\
R_{ jr }^{ ir }&=&\frac { 1 }{ N }\, {\partial }_{ r }{K }_{ j }^{ i}-{K }_{ n }^{ i }{K }_{ j }^{ n }\,,  \nonumber \\
{R }_{ jl }^{ ik }&=&{\mathcal{R }}_{ jl }^{ ik }(h)-{K }_{ j }^{ i }{K }_{ l }^{ k }+{K }_{ l }^{ i }{K }_{ j }^{ k }\,.
\end{eqnarray}
Here, $\nabla_{j}=\nabla_{j}(h)$ is the covariant derivative defined with respect to the induced metric and ${\mathcal{R }}_{ jl }^{ ik}(h)$ is the intrinsic curvature of the boundary $\partial \mathcal{M}$.

\section{Explicit derivation of Conformal Mass in Lovelock AdS gravity}
\label{truncationapp}

In this Appendix, we provide further details regarding the factorization and the truncation of the Kounterterm charges in even and odd dimensions, in order to obtain an expression for Conformal Mass. We will employ a shorthand where $\delta^{[k] }$ represents the antisymmetric Kronecker
delta of rank $k$, that is, $\delta _{i_1\cdots i_k}^{j_1\cdots j_k}$. In turn, $\delta ^{j[k]}_{i[k]}$ denotes a totally antisymmetric Kronecker delta of rank $(k+1)$ with the indices $i$ and $j$ fixed, that is, $\delta _{ii_1\cdots i_k}^{jj_1\cdots j_k}$.

\subsection{Even dimensions}

The charge density tensor in even dimensions ($D=2n$) is given by Eq.(\ref{qeven}). In order to manipulate it as a polynomial of the spacetime curvature, we employ the shorthand just defined above,
\begin{eqnarray}
\tau_{i}^{j} &=&\frac{1}{16\pi G}\,\delta ^{j[
2n-2]}_{m[2n-2]}\,K^{m}_{i}\sum_{p=1}^{n-1}{\frac{p{\alpha }_{p}}{{2}^{n-2}(2n-2p)!}}\left[ R^{p-1}\left(\delta ^{[2]}\right)^{n-p}-\left( -{{\ell _{%
\mathrm{eff}}^{2}}}\right) ^{n-p}R^{n-1}\right]  \nonumber \\
&=&\frac{{{\ell _{\mathrm{eff}}^{2n-2}}}}{16\pi G{2}^{n-2}}\,\delta ^{j[
2n-2]}_{m[2n-2]}\,K^{m}_{i}\,\sum_{p=1}^{n-1}{\frac{p{\alpha }_{p}}{\left(
2n-2p\right) !}}\left( \frac{R}{{{\ell _{\mathrm{eff}}^{2}}}}\right) ^{p-1}%
\left[ \left( \frac{\delta ^{\lbrack
2]}}{{{\ell _{\mathrm{eff}}^{2}}}}\,\right) ^{n-p}-( -R) ^{n-p}\right].
\end{eqnarray}
The polynomial in $R$ can conveniently be factorized using the following identity,
\begin{equation}
{a}^{n-p}-{b}^{n-p}=(n-p)(a-b)\int\limits_{0}^{1}{ds\,{\left[ s(a-b)+b\right]}^{n-p-1},}  \label{difference}
\end{equation}
with $a=\frac{1}{{\ell _{\mathrm{eff}}^{2}}}\delta ^{[2]}$ and $b=R$, what yields
\begin{eqnarray}
\left( \frac{1}{{\ell _{\mathrm{eff}}^{2}}}\,\delta ^{\lbrack 2]}\right)
^{n-p}-{(-R)}^{n-p} &=&(n-p)\,F\int\limits_{0}^{1}{ds\,}\left[ {s}\left( R+%
\frac{1}{{\ell _{\mathrm{eff}}^{2}}}\,\delta ^{\lbrack 2]}\right) {-R}\right]^{n-p-1}.
\end{eqnarray}
The AdS curvature $F$ is defined as in Eq.(\ref{F}). The charge density tensor adopts the
factorized form,
\begin{equation}
\tau_{i}^{j}=\frac{{{\ell _{\mathrm{eff}}^{2n-2}}}}{16\pi G}\,{\delta }_{{i}%
_{1}\dots {i}_{2n-1}}^{j{j}_{2}\dots {j}_{2n-1}}{K}_{i}^{{i}_{1}}
F_{{j}_{2}{j}_{3}}^{{i}_{2}{i}_{3}}\mathcal{P}_{j_4\dots j_{2n-1}}^{i_4\dots i_{2n-1}}(R)\,,
\end{equation}
where the polynomial is given by
\begin{equation}
\mathcal{P}(R)=\sum_{p=1}^{n-1}{\frac{p{\alpha }_{p}}{{2}^{n-1}\left( 2n-2p-1\right) !}}%
\left( \frac{R}{{{\ell _{\mathrm{eff}}^{2}}}}\right)
^{p-1}\int\limits_{0}^{1}{ds\,}\left[ {s}\left( R+\frac{1}{{\ell _{\mathrm{%
eff}}^{2}}}\,\delta ^{\lbrack 2]}\right) -{R}\right] ^{n-p-1}.  \label{peven}
\end{equation}

Next, one notices that the asymptotic behavior of the AdS curvature is such that
one can use $W$ instead of $F$ in the above formula
\begin{equation}
\tau_i^j=\frac{\ell_{\mathrm{eff}}^{2n-2}}{16\pi G}\,\delta_{{i}%
_{1}{i}_{2}\dots {i}_{2n-1}}^{j{j}_{2}\dots {j}_{2n-1}}{K}_{i}^{{i}_{1}}\, W_{{j}_{2}{j}_{3}}^{{i}_{2}{i}_{3}} \mathcal{P}_{j_4\dots j_{2n-1}}^{i_4\dots i_{2n-1}}(R)+\mathcal{O}(1/r^{2D-2})\,.
\end{equation}
In addition, one considers the asymptotic expansion of the relevant tensors (\ref{KR}), that is,
${K}_{j}^{i}=-\frac{1}{\ell _{\mathrm{eff}}}\,{\delta }_{j}^{i}+\mathcal{O}(1/r)$ and
$R=-\frac{1}{{\ell _{\mathrm{eff}}^{2}}}\delta ^{[2]}+\mathcal{O}(1/r^{D-1})$, what leads to
\begin{equation}
\tau_i^j = -\frac{\ell_{\mathrm{eff}}^{2n-3}}{16\pi G}\,{ \delta  }_{ i{ i }_{ 2 }\dots { i }_{ 2n-1 } }^{ j{ j }_{ 2 }\dots { j }_{ 2n-1 } }\,W_{{j}_{2}{j}_{3}}^{{i}_{2}{i}_{3}}\,\mathcal{P}_{j_4\dots j_{2n-1}}^{i_4\dots i_{2n-1}}(-\ell_{\mathrm{eff}}^{-2}\,\delta^{[2]}) +{\mathcal{O}}(1/{r}^{2D-2})\,,
\end{equation}
where the polynomial is evaluated at the leading order in the curvature $R=-\ell_{\mathrm{eff}}^{-2}\,\delta^{[2]}$. A simple computation produces
\begin{eqnarray}
\mathcal{P}(-\ell_{\mathrm{eff}}^{-2}\,\delta^{[2]}) &=&\sum_{p=1}^{n-1}{\frac{p(n-p){\alpha }_{p}}{{2}%
^{n-2}\left( 2n-2p\right) !}}\left( \frac{1}{{{\ell _{\mathrm{eff}}^{2}}}}%
\right) ^{p-1}\left( -\frac{1}{{\ell _{\mathrm{eff}}^{2}}}\,\delta ^{[2]}\right) ^{p-1}\int\limits_{0}^{1}{ds\,}\left( \frac{1}{{\ell _{\mathrm{eff%
}}^{2}}}\,\delta ^{\lbrack 2]}\right) {^{n-p-1}} \nonumber \\
&=&\frac{\Delta ^{\prime }(\ell _{\mathrm{eff}}^{-2})}{\left( 2n-3\right) !\,%
{2}^{n-1}{\ell _{\mathrm{eff}}^{2(n-2)}}}\,\left(\delta ^{[2]}\right)^{n-2}\,.
\end{eqnarray}
When plugged in the charge formula, one obtains
\begin{eqnarray}
\delta_{{ii}_{2}\cdots {i}_{2n-1}}^{j{j}_{2}\cdots {j}_{2n-1}}\,
\mathcal{P}_{{j}_{4}\cdots {j}_{2n-1}}^{{i}_{4}\cdots {i}_{2n-1}}(-\ell_{\mathrm{eff}}^{-2}\,\delta^{[2]}) =\frac{\Delta'(\ell _{\mathrm{eff}}^{-2})}{2\left( 2n-3\right) \,
{\ell _{\mathrm{eff}}^{2(n-2)}}}\,{\delta }_{{ii}_{2}{i}_{3}}^{j{j}_{2}j_{3}}\,.
\end{eqnarray}
The charge density takes the form

\begin{equation}
\tau_{i}^{j} =-\frac{{\ell _{\mathrm{eff}}}}{16\pi G}\,\frac{\Delta'(\ell _{\mathrm{eff}}^{-2})}{2\left( 2n-3\right) }\,{\delta }_{{ii}_{2}{i}_{3}}^{j{j}_{2}j_{3}}\,{{W}_{{j}_{2}{j}_{3}}^{{i}_{2}{i}_{3}}}+\mathcal{O}(1/r^{2D-2})\,,
\end{equation}
what was shown to be proportional to the electric part of the Weyl tensor in Subsection \ref{even_dim}.

\subsection{Odd dimensions}
\label{oddapp}

The charge density tensor (\ref{qodd}) in odd dimensions ($D=2n+1$) can be symbolically written as
\begin{equation}
\tau_{i}^{j} =\frac{\delta_{m[2n-1]} ^{j[2n-1]}}{16\pi G2^{n-2}}\,K^{m}_{i} \left[ \sum_{p=1}^{n}\frac{p\alpha _{p}}{(2n-2p+1)!}\,R^{p-1}\left(\delta ^{\lbrack 2]}\right)^{n-p}+16\pi G\,nc_{2n}\int\limits_{0}^{1}du\,\left( R+\frac{u^{2}}{\ell _{\textrm{eff}}^{2}}\,\delta ^{[2]}\right) ^{n-1}\right]  \label{qodd0}
\end{equation}
where the coupling constant $c_{2n}$ is given by Eq.(\ref{c_odd}). Introducing the constant
\begin{equation}
\gamma = \sum_{p=1}^{n}\frac{(-1)^{p}p\alpha _{p}\,\ell _{\textrm{eff}}^{2(n-p)}}{(2n-2p+1)!}\,,
\end{equation}
the tensor $\tau_{i}^{j}$ can be cast into the form
\begin{equation}
\tau_{i}^{j}=\frac{nc_{2n}}{2^{n-2}\gamma }\,\delta_{m[2n-1]} ^{j[2n-1]}K^{m}_{i}
\sum_{p=1}^{n}\frac{(-1)^{p}p\alpha _{p}\,\ell _{\textrm{eff}}^{2(n-p)}}{(2n-2p+1)!}\int\limits_{0}^{1}du\,\mathcal{I}_{p}(u)\,.
\end{equation}
Here, the tensorial quantity $\mathcal{I}_{p}(u)$ is defined as
\begin{equation}
\mathcal{I}_{p}(u)=\left( R+\frac{u^{2}}{\ell _{\textrm{eff}}^{2}}\,\delta ^{\lbrack 2]}\right) ^{n-1}-\left( u^{2}-1\right) ^{n-1}
(-R)^{p-1}\left(\frac{1}{\ell _{\textrm{eff}}^{2}}\,\delta
^{\lbrack 2]}\right) ^{n-p}.
\end{equation}
In order to factorize $\tau^j_i$ by the AdS curvature $F$, one may add zero in the following way
\begin{equation}
-(-R)^{p-1}=-\left( \frac{1}{\ell _{\textrm{eff}}^{2}}
\delta ^{[2]}\right) ^{p-1}+\left( \frac{1}{\ell _{\textrm{eff}}^{2}}\,\delta ^{[2]}\right) ^{p-1}-(-R) ^{p-1}
\label{decomposition}
\end{equation}
and use once again the formula (\ref{difference}) in the last two
terms of the last expression with $a=\frac{1}{\ell _{\textrm{eff}}^{2}}\,\delta ^{\lbrack 2]}$,
$b=R$ and $k=p-1$. In doing so, one obtains the relation
\begin{equation}
-(-R)^{p-1}=-\left( \frac{1}{\ell _{\textrm{eff}}^{2}}
\,\delta ^{\lbrack 2]}\right) ^{p-1}+\left( p-1\right) F\int\limits_{0}^{1}{ds{\left[ s\left(R+\frac{1}{\ell _{\textrm{eff}}^{2}}\,\delta ^{[2]}\right)-R\right] }^{p-2}}
\end{equation}
which, when substituted  in $\mathcal{I}_{p}$, produces
\begin{eqnarray}
\mathcal{I}_{p} &=&\left( R+\frac{u^{2}}{\ell _{\textrm{eff}}^{2}}
\,\delta ^{\lbrack 2]}\right) ^{n-1}-\left( u^{2}-1\right) ^{n-1}\left(
\frac{1}{\ell _{\textrm{eff}}^{2}}\,\delta ^{\lbrack 2]}\right) ^{n-1}\nonumber
\\
&&+F\left( p-1\right) \left( u^{2}-1\right) ^{n-1}
\left( \frac{1}{\ell _{\textrm{eff}}^{2}}\,\delta ^{\lbrack 2]}\right)^{n-p}
\int\limits_{0}^{1}{ds\left[ \left( {s-1}\right)
{{F+\frac{1}{\ell _{\textrm{eff}}^{2}}\,\delta ^{\lbrack 2]}}}\right] ^{p-2}.}
\end{eqnarray}
Applying the formula (\ref{difference}) yet another time in the first two terms of Eq.(\ref{decomposition}), where now
$a=R+\frac{u^{2}}{\ell _{\textrm{eff}}^{2}}\,\delta ^{[2]}$, $b=-(u^{2}-1) \frac{1}{\ell _{\textrm{eff}}^{2}}\,
\delta ^{[2]}$ and $k=n-1$, leads to a factorization of $\mathcal{I}_{p} $,
\begin{eqnarray}
\mathcal{I}_{p} &=&\left( n-1\right) F\int\limits_{0}^{1}{ds}
\left( {{sF+\frac{u^{2}-1}{\ell _{\textrm{eff}}^{2}}\,\delta ^{\lbrack 2]}}}\right) ^{n-2} \nonumber\\
&&+(p-1)F\left(\frac{1}{\ell_{\textrm{eff}}^2}\,\delta ^{[2]}\right)^{n-p}(u^2-1)^{n-1}
\int\limits_0^1 ds\left[(s-1)F+\frac{1}{\ell_{\textrm{eff}}^2}\,\delta^{[2]}\right]^{p-2}\,.
\end{eqnarray}
In that way, the charge density tensor is factorized, as well,
\begin{equation}
\tau_{i}^{j}=\frac{{{\ell _{\mathrm{eff}}^{2n-2}}}}{16\pi G}\,{\delta }_{{i}%
_{1}{i}_{2}\dots {i}_{2n-1}}^{j{j}_{2}\dots {j}_{2n-1}}{K}_{i}^{{i}_{1}}
F_{{j}_{2}{j}_{3}}^{{i}_{2}{i}_{3}}\tilde{\mathcal{P}}_{j_4\dots j_{2n-1}}^{i_4\dots i_{2n-1}}(R)\,,
\end{equation}
where $\tilde{\mathcal{P}}$ is a polynomial of the type

\begin{eqnarray}
\label{tildeP}
\tilde{\mathcal{P}}=-\frac{1}{2^{n-2}}\left[ \int\limits_{ 0 }^{ 1 } du (u^2-1)^{n-1}  \right]^{ -1 }\sum_{p=1}^{n}\frac{p\alpha _{p}\left(-\ell_{\mathrm{eff}} ^{2}\right)^{-(p-1)}}{(2n-2p+1)!}\,\mathcal{J}_{p}(R)\,,
\end{eqnarray}
with the help of the parametric integrals
\begin{eqnarray}
\mathcal{J}_{p}(R) &=&(n-1) \int\limits_{0}^{1}du\int\limits_0^1 ds \left( sR+\frac{u^2+s-1}{\ell_{\textrm{eff}}^2}\,\delta ^{[2]}\right)^{n-2}  \nonumber
\\
&&+(p-1)\left( \frac{1}{\ell _{\textrm{eff}}^2}\,\delta ^{[2]}\right)^{n-p}
\int\limits_{0}^{1}du\,(u^2-1)^{n-1}\int\limits_{0}^{1}{ds\left[(s-1)R+\frac{s}{\ell _{\textrm{eff}}^2}\,\delta ^{[2]}\right]^{p-2}}
\label{int}
\end{eqnarray}

In a similar manner as in the even-dimensional case, there is no change in the energy of a Lovelock black hole if one replaces the AdS curvature by the Weyl tensor in the expression for $\tau_{i}^{j}$, i.e.,
\begin{equation}
\tau_{i}^{j}=\frac{{{\ell _{\mathrm{eff}}^{2n-2}}}}{16\pi G}\,{\delta }_{{i}%
_{1}{i}_{2}\dots {i}_{2n-1}}^{j{j}_{2}\dots {j}_{2n-1}}{K}_{i}^{{i}_{1}}
W_{{j}_{2}{j}_{3}}^{{i}_{2}{i}_{3}}\tilde{\mathcal{P}}_{j_4\dots j_{2n-1}}^{i_4\dots i_{2n-1}}(R)\,.
\end{equation}
Because the Weyl tensor is already of order $\mathcal{O}(1/r^{D-1})$, we can drop subleading contributions in $K$, $R$ and the polynomial
$\tilde{\mathcal{P}}$

\begin{equation}
\tau_{i}^{j}=-\frac{{{\ell _{\mathrm{eff}}^{2n-3}}}}{16\pi G}\,{\delta }_{{i}%
{i}_{2}\dots {i}_{2n-1}}^{j{j}_{2}\dots {j}_{2n-1}}
W_{{j}_{2}{j}_{3}}^{{i}_{2}{i}_{3}}\tilde{\mathcal{P}}_{j_4\dots j_{2n-1}}^{i_4\dots i_{2n-1}}(-\ell_{\mathrm{eff}}^{-2}\,\delta^{[2]})\,.
\end{equation}

 In other words, it is sufficient to evaluate $\mathcal{J}_{p}$ in Eq.(\ref{int}) for global AdS space.
For that value of the curvature, the integral in $s$ turns trivial, such that
\begin{equation}
\mathcal{J}_{p}(-\ell _{\textrm{eff}}^{-2}\delta^{[2]})=\left( \frac{1}{\ell _{\textrm{eff}}^{2}}\,\delta^{[2]}\right) ^{n-2}
\int\limits_{0}^{1}du\left[ \left( n-1\right) {{\left( u^{2}-1\right) ^{n-2}}}+\left( p-1\right) \left( u^{2}-1\right) ^{n-1}\right] .
\label{int1}
\end{equation}
In turn, the integral in $u$ becomes
\begin{equation}
\int\limits_{0}^{1}du\,\left( u^{2}-1\right) ^{n-2}=-\frac{2n-1}{2(n-1) }\int\limits_{0}^{1}du\,(u^{2}-1)^{n-1},
\end{equation}
what finally can be expressed as
\begin{equation}
\mathcal{J}_{p}({-\ell_{\textrm{eff}}^{-2}\delta^{[2]}})=-\frac{2n-2p+1}{2}\left( \frac{1}{\ell _{\textrm{eff}}^{2}}\,\delta ^{\lbrack 2]}\right) ^{n-2}\int\limits_{0}^{1}du\,\left(u^{2}-1\right) ^{n-1}.
\end{equation}
Substituting this result into the polynomial yields

\begin{eqnarray}
\tilde{\mathcal{P}}&=&\frac{\left(-\ell_{\textrm{eff}}^{2}\right)^{-n+2}}{2^{n-1}} \left(\delta ^{[2]}\right)^{n-2} \sum_{p=1}^{n}\frac{(-1)^{p-1}p\alpha_{p}}{(2n-2p)!}\left(\frac{1}{\ell_{\mathrm{eff}} ^{2}}\right)^{p-1}\, \nonumber \\
&=& \frac{\Delta ^{\prime }(\ell _{\mathrm{eff}}^{-2})}{\left(2n-2\right)!2^{n-1}\ell_{\textrm{eff}}^{2\left(n-2\right)}}\left(\delta ^{[2]}\right) ^{n-2}.
\end{eqnarray}
Then, the linearization of the charge density leads to something proportional to the Weyl tensor
\begin{eqnarray}
\tau_{i}^{j} =-\frac{{\ell _{\mathrm{eff}}}}{16\pi G}\,\frac{\Delta'(\ell _{\mathrm{eff}}^{-2})}{2\left( 2n-2\right) }\,{\delta }_{{ii}_{2}{i}_{3}}^{j{j}_{2}j_{3}}\,{{W}_{{j}_{2}{j}_{3}}^{{i}_{2}{i}_{3}}}+\mathcal{O}(1/r^{2D-2})\,.
\end{eqnarray}

\bibliographystyle{JHEP}
\bibliography{refjhep}

\end{document}